\documentclass[reprint, aps,prl,twocolumn,]{revtex4-1}
\usepackage{amssymb}
\usepackage{amsmath}
\usepackage{epsfig}
\usepackage{color}
\usepackage{graphics, graphicx}
\usepackage{bbold}
\usepackage{psfrag}
\usepackage{mathcomp}
\usepackage{subfigure}
\usepackage{verbatim}
\usepackage[colorlinks, citecolor=blue]{hyperref}
\usepackage[normalem]{ulem}
\begin{document}

\title{Spin-orbit coupling induced quantum droplet in ultracold Bose-Fermi mixtures}
\author{Xiaoling Cui}
%\email{xlcui@iphy.ac.cn} 
\affiliation{Beijing National Laboratory
for Condensed Matter Physics, Institute of Physics, Chinese Academy
of Sciences, Beijing, 100190, People's Republic of China}
\date{\today}

\begin{abstract}
Quantum droplets have intrigued much attention recently in view of their successful observations in the ultracold homonuclear atoms.  In this work, we demonstrate a new mechanism for the formation of quantum droplet in heteronuclear atomic systems, i.e., by applying the synthetic spin-orbit coupling(SOC). Take the Bose-Fermi mixture for example, we show that by imposing a Rashba SOC between the spin states of fermions, the greatly suppressed Fermi pressure can enable the formation of Bose-Fermi droplets even for very weak boson-fermion attractions, which are insufficient to bound a droplet if without SOC.  In such SOC-induced quantum droplets, the boson/fermion density ratio universally depends on the SOC strength, and they occur in the mean-field collapsing regime but with a negative fluctuation energy, distinct from the interaction-induced droplets found in literature. The accessibility of these Bose-Fermi droplets in ultracold Cs-Li and Rb-K mixtures is also discussed. Our results shed light on the droplet formation in a  vast class of heteronuclear atomic systems through the manipulation of single-particle physics. 
\end{abstract}
\maketitle

%{\it Introduction.}
Self-bound droplets are ubiquitous in nature, while their quantum mechanical analogs, quantum droplets, are challenging to achieve in physical systems as their appearance requires sophisticated balance between attractive and repulsive forces. %and  are not that obvious to achieve . 
% definition, is a self-bound object stabilized by quantum mechanical effect. %In general, both attractive and repulsive forces are required in its formation, to make it self-bound while avoiding collapse. 
Recently, the study of quantum droplet has become a hot topic in the field of ultracold atoms. A pioneer work by Petrov showed that self-bound droplets of a two-component Bose gas can form in the mean-field collapsing regime\cite{Petrov}, due to a balance between mean-field attraction ($\sim -n^2$, $n$ is the density) and Lee-Huang-Yang repulsion from quantum fluctuations ($\sim n^{5/2}$), and importantly, they stay in the weak coupling regime that can effectively avoid atom loss. To date, quantum droplets have been successfully observed in Lanthanum atoms with strong dipole-dipole interaction\cite{Pfau_1,Pfau_2,Pfau_3,Ferlaino}, and in alkali boson mixtures\cite{Tarruell_1,Tarruell_2,Inguscio} which exactly follow Petrov's scenario. Droplet formation has recently also been predicted in low-D\cite{Petrov_2, Santos, Nishida,Petrov_3} and in photonic systems\cite{Valiente}.

Given successful explorations of quantum droplets in homonuclear systems\cite{Pfau_1,Pfau_2,Pfau_3,Ferlaino,Tarruell_1,Tarruell_2,Inguscio}, it naturally arises a question whether such a peculiar state exists in heteronuclear systems, especially Bose-Fermi mixtures with coexisting different statistics. %Answering this question will further deepen our understanding of the general principle for , 
%On a first sight, there seems little difference between Bose-Fermi and Bose-Bose systems, as the Fermi pressure in the former naturally serves as a repulsive force, similar to the role of boson repulsion in the latter. 
Actually, in this problem the Bose-Fermi and Bose-Bose mixtures share some similarities, in that the Fermi pressure in the former naturally plays the role of boson repulsion in the latter as a repulsive force, and both systems host an additional repulsion from quantum fluctuations\cite{Petrov, Giorgini}.  So a Bose-Fermi droplet is expectable by fine-tuning boson-fermion attractions, as has been theoretically confirmed recently\cite{droplet_BF}. Nevertheless, one notes that the Fermi pressure scales as $\sim n^{5/3}$, which, compared to the boson repulsion ($\sim n^2$), generates higher repulsive force in the dilute limit. Accordingly, a stronger attraction in Bose-Fermi mixtures is required to form a droplet.  Strong interaction can invalidate perturbative theories in treating the droplets, and inevitably induce severe atom losses to prevent their realistic detection in experiments. 
% . It is thus imperative to discover a new mechanism to ensure the droplet formation in such system.

In this work, we demonstrate a new route to stabilize the quantum droplet, i.e.,  by introducing the spin-orbit coupling (SOC). In the past few years, cold atoms experiments have successfully realized the synthetic 1D\cite{Spielman_exp1,Spielman_exp2,Shuai,Spielman_exp3,Jing,MIT,Chuanwei,Spielman_exp4,Shuai_2013,Spielman_2013,Jing_2013} and 2D\cite{2Dsoc_Jing,2Dsoc_Jing_2,2Dsoc_Shuai} types of SOC, and the highly symmetric SOC including the Rashba and isotropic types have also been theoretically proposed\cite{Rashba_Spielman_1, Rashba_Spielman_2, Rashba_Spielman_3, Rashba_Xu_1, Rashba_Xu_2, Rashba_Liu,spielman_3d_1,spielman_3d_2}. %There have also been a number of proposals to realize highly . 
Our work is simply motivated by the fact that SOCs can significantly modify the single-particle physics in low-energy space. In particular, for a highly symmetric SOC, %the low-energy density of state(DoS) can be greatly enhanced due to 
the resulted single-particle ground state degeneracy in combination with interactions has been found to induce intriguing dimer\cite{Vijay, Cui, ZhangPeng, Yu, Greene, Blume}, trimer\cite{Shi_1,Shi_2,Cui_Yi} and many-body physics\cite{review}. Here, we point out another dramatic effect of the highly symmetric SOC, namely, in driving the formation of stable Bose-Fermi droplets in weak coupling regime. The associated mechanism can be generalized to various other heteronuclear systems in different dimensions.% a purely SOC-induced Boseit can  show that the highly symmetric SOC can facilitate the droplet formation in Bose-Fermi mixtures through the effective suppression of Fermi pressure of a non-interacting Fermi gas, thus facilitating the droplet formation in Bose-Fermi mixtures. 

To be concrete, we consider a Rashba spin-orbit coupled Fermi gas spin-selectively interacting with a Bose gas, which is described by the following  Hamiltonian
\begin{eqnarray}
H&=&\sum_{\bf k}\epsilon_{\bf k}^b b^{\dag}_{\bf k}b_{\bf k} + \frac{U_{bb}}{V}\sum_{{\bf k,k',Q}}b^{\dag}_{{\bf k}}b^{\dag}_{\bf Q-k}b_{\bf Q-k'}b_{{\bf k'}}\nonumber\\
&+&\sum_{{\bf k},\alpha}\epsilon_{\bf k}^f f^{\dag}_{{\bf k},\alpha}f_{{\bf k},\alpha}+\frac{\lambda}{m_f} \sum_{\bf k}\left( (k_x-ik_y) f^{\dag}_{{\bf k},\uparrow}f_{{\bf k},\downarrow} + h.c. \right) \nonumber\\
&+&\frac{U_{bf}}{V}\sum_{{\bf k,k',Q}}f^{\dag}_{{\bf k},\uparrow}b^{\dag}_{\bf Q-k}b_{\bf Q-k'}f_{{\bf k'},\uparrow}.  \label{H}
\end{eqnarray}
Here $b^{\dag}_{\bf k}$ and $f^{\dag}_{{\bf k},\alpha}$ create a boson and a spin-$\alpha$($=\uparrow,\downarrow$) fermion, respectively, with energy $\epsilon_{\bf k}^b={\bf k}^2/2m_b$ and  $\epsilon_{\bf k}^f={\bf k}^2/2m_f$;  $U_{bb}$ and $U_{bf}$ are respectively the bare boson-boson and boson-fermion interactions, which can be related to scattering lengths $a_{bb}$ and $a_{bf}$ via  renormalization equations, for instance, $1/U_{bf}=1/g_{bf}-(1/V)\sum_{\bf k} 1/(2m_{bf} {\bf k}^2)$, with $g_{bf}=2\pi a_{bf}/m_{bf}$, $m_{bf}=m_bm_f/(m_b+m_f)$, and $V$ the volume. 
Here we consider a Rashba SOC between two-species fermions with strength $\lambda$, and the resulted single-particle eigenstate is created by $f^{\dag}_{{\bf k},\sigma}=\sum_{\alpha} \gamma^{\alpha}_{{\bf k},\sigma} f^{\dag}_{{\bf k},\alpha}$, where $\sigma=\pm$ is the index of helicity branch, $\gamma^{\uparrow}_{{\bf k},\pm}=\pm e^{\pm i\phi_{k}/2}/\sqrt{2},\ \gamma^{\downarrow}_{{\bf k},\pm}= e^{\pm i\phi_{k}/2}/\sqrt{2}, \phi_k={\rm arg}(k_x,k_y)$; the corresponding eigen-energy is $\epsilon^f_{{\bf k},\sigma}=\left( (k_{\perp}+\sigma\lambda)^2+k_z^2\right)/(2m_f)$ (here $k_{\perp}=\sqrt{k_x^2+k_y^2}$), which gives a $U(1)$ ground state degeneracy in ${\bf k}$-space with $k_{\perp}=\lambda$. For brevity, we  take $\hbar=1$ throughout the paper.  

In this work, we consider weakly interacting bosons with small $a_{bb}(>0)$, and a weak attraction between boson and spin-$\uparrow$ fermion with small $a_{bf}(<0)$. Given the boson and fermion densities $n_b$ and $n_f$, the energy density of the system can be written as 
\begin{equation}
{\cal E}(n_b,n_f)={\cal E}_b+{\cal E}_f+{\cal E}_{bf}, \label{E}
\end{equation}
here ${\cal E}_b=(2\pi a_{bb}/m_b)n_b^2[1+(128/15\pi^{1/2})(n_ba_{bb}^3)^{1/2}]$ is the ground state energy of the Bose gas with Lee-Huang-Yang correction. ${\cal E}_f$ is the Fermi sea energy under Rashba SOC:
\begin{equation}
{\cal E}_f=\frac{1}{V}\sum_{{\bf k},\sigma}\epsilon^f_{{\bf k},\sigma}\theta(E_f-\epsilon^f_{{\bf k},\sigma}),
\end{equation}
with $E_f\equiv \lambda_f^2/(2m_f)$ the Fermi energy and $\lambda_f$ the Fermi momentum, determined by the density constraint $n_f=\frac{1}{V}\sum_{{\bf k},\sigma}\theta(E_f-\epsilon^f_{{\bf k},\sigma})$. ${\cal E}_{bf}={\cal E}_{bf}^{(1)}+{\cal E}_{bf}^{(2)}$ is the interaction energy between bosons and fermions, %which is composed by mean-field energy $\sim g_{bf}n_bn_{f,\uparrow}$ and the beyond-mean-field ones due to quantum fluctuations. Here 
here ${\cal E}_{bf}^{(1)}= g_{bf}n_bn_{f,\uparrow}$ is the mean-field interaction energy, and ${\cal E}_{bf}^{(2)}\ (\sim g_{bf}^2)$ is the lowest-order correction due to density fluctuations, which can be obtained from the second-order perturbation theory as:
%to take into account the virtual excitation of a single Bogoliubov quasiparticle of bosons together with one particle-hole excitation of fermions out of the spin-orbit coupled Fermi sea, which finally results in
%\begin{eqnarray}
%{\cal E}_{bf}={\cal E}_{bf}^{(1)}+{\cal E}_{bf}^{(2)},
%{\cal E}_{bf}&=&n_bn_{f,\uparrow}(g_{bf}+\frac{g_{bf}^2}{V}\sum_{\bf k}\frac{2m_{bf}}{{\bf k}^2})\nonumber\\
%&&-n_b\frac{g_{bf}^2}{V^2}\sum_{\bf k} \frac{\epsilon_{\bf k}^b}{\omega_{\bf k}}\sum_{{\bf q},\sigma,\sigma'} \frac{1}{4}\frac{\theta(E_f-\epsilon_{{\bf q},\sigma}^f)\theta(\epsilon_{{\bf k+q},\sigma'}^f-E_f)}{\omega_k+\epsilon_{{\bf k+q},\sigma'}^f-\epsilon_{{\bf q},\sigma}^f}.
%\end{eqnarray}
\begin{eqnarray}
{\cal E}_{bf}^{(2)}&=&n_b\frac{g_{bf}^2}{V} \sum_{\bf k}\left( n_{f,\uparrow}\frac{2m_{bf}}{{\bf k}^2}   \right. \nonumber\\
&&-\left. \frac{\epsilon_{\bf k}^b}{\omega_{\bf k}}\sum_{{\bf q},\sigma,\sigma'} \frac{1}{4V}\frac{\theta(E_f-\epsilon_{{\bf q},\sigma}^f)\theta(\epsilon_{{\bf k+q},\sigma'}^f-E_f)}{\omega_k+\epsilon_{{\bf k+q},\sigma'}^f-\epsilon_{{\bf q},\sigma}^f} \right). \label{Ebf2}
\end{eqnarray}
Here $\omega_{\bf k}=\sqrt{\epsilon_{\bf k}^b(\epsilon_{\bf k}^b+8\pi n_ba_{bb}/m_b)}$ is the Bogoliubov excitation energy of bosons. %In obtaining ${\cal E}_{bf}^{(2)}$, we have taken into account the virtual excitation of a single Bogoliubov quasiparticle of bosons together with one particle-hole excitation of fermions out of the spin-orbit coupled Fermi sea. 
In the limit of $\lambda\rightarrow 0$, our result recovers the perturbative energy of Bose-Fermi mixtures without SOC\cite{Giorgini}. 
%The second-order perturbation treatment was previously applied to Bose-Fermi mixture without SOC\cite{Giorgini,Wilkens}, and here our results can recover the result of Ref.\cite{Giorgini} if send $\lambda$ to zero. 
%We will show that the introduction of SOC will 

Given ${\cal E}(n_b,n_f)$ in (\ref{E}), one can obtain the chemical potentials $\mu_{b}=\partial{\cal E}/\partial n_{b},\ \mu_{f}=\partial{\cal E}/\partial n_{f}$, and the pressure density ${\cal P}={\cal P}_b+{\cal P}_f+{\cal P}_{bf}$, where
\begin{eqnarray}
{\cal P}_b&=&n_b\frac{\partial {\cal E}_b}{\partial n_b} - {\cal E}_b;\ \ \ \ {\cal P}_f=n_f\frac{\partial {\cal E}_f}{\partial n_f} - {\cal E}_f;\\
{\cal P}_{bf}&=& {\cal P}_{bf}^{(1)}+{\cal P}_{bf}^{(2)};\nonumber\\
{\cal P}_{bf}^{(i)} &=& n_b\frac{\partial {\cal E}_{bf}^{(i)}}{\partial n_b}+n_f\frac{\partial {\cal E}_{bf}^{(i)}}{\partial n_f}-{\cal E}_{bf}^{(i)}  \ \ (i=1,2);
\end{eqnarray}
Here ${\cal P}_{b}$ (${\cal P}_{f}$) is the pressure caused by individual bosons (fermions), ${\cal P}_{bf}$ is due to boson-fermion interactions and contributed from both the mean-field (${\cal P}_{bf}^{(1)}$) and the quantum  fluctuation (${\cal P}_{bf}^{(2)}$) parts. The introduction of SOC will not change ${\cal P}_b$ and ${\cal P}_{bf}^{(1)}$, but will strongly modify ${\cal P}_f$ and ${\cal P}_{bf}^{(2)}$ as shown below.

Before proceeding, we should note that a stable ground state droplet occurs when the following conditions are simultaneously satisfied:
\begin{eqnarray}
{\rm (i)}\  && {\cal E}<0,\ \ {\cal P}=0;  \nonumber \\
{\rm (ii)} \  && \mu_b\frac{ \partial{\cal P}}{\partial n_{f}}= \mu_f\frac{ \partial{\cal P}}{\partial n_{b}};\label{E_min} \nonumber\\
{\rm (iii)} \  && \frac{ \partial\mu_b}{\partial n_{b}}>0,\ \frac{ \partial\mu_f}{\partial n_{f}}>0,\ \frac{ \partial\mu_b}{\partial n_{b}}\frac{ \partial\mu_f}{\partial n_{f}}>(\frac{ \partial\mu_b}{\partial n_{f}})^2 \label{stable_condition} \nonumber
\end{eqnarray}
here the condition (i) describes a self-bound object that is in equilibrium with vacuum, a characteristic feature of droplet\cite{Petrov}; condition (ii) further searches for the ground state droplet with minimal energy\cite{droplet_BF}; and (iii) ensures the droplet be stable against density fluctuations. 

\begin{figure}[h]
\includegraphics[height=4.8cm]{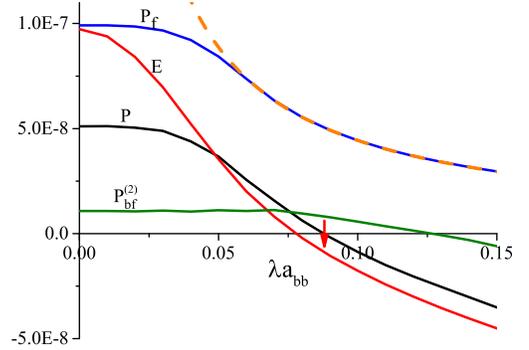}
\caption{${\cal E},\ {\cal P}$, ${\cal P}_f$ and ${\cal P}_{bf}^{(2)}$ [in units of $m_ba_{bb}^5/(2\pi V)$] as functions of $\lambda a_{bb}$. Here we take $n_ba_{bb}^3=2\times 10^{-5}$, $n_fa_{bb}^3=10^{-4}$, $a_{bf}=-3a_{bb}$, and $m_b/m_f=133/6$. The red arrow marks the location where the droplet condition (i) is satisfied. The orange dashed line shows fit to ${\cal P}_f$ according to Eq.\ref{P_rashba}.} \label{fig1}
\end{figure}

To gain the first insight on how a Rashba SOC affect the droplet formation, in Fig.\ref{fig1} we take $^{133}$Cs-$^{6}$Li system and show its energy ${\cal E}$, pressure ${\cal P}$ and pressure components ${\cal P}_f,\ {\cal P}_{bf}^{(2)}$ as functions of SOC strength $\lambda$, for a given attraction $a_{bf}=-3a_{bb}$ and given densities $n_ba_{bb}^3=2\times 10^{-5}$, $n_fa_{bb}^3=10^{-4}$. It can be seen that as increasing $\lambda$ from zero, both ${\cal E}$ and ${\cal P}$ decrease monotonically, such that at a critical $\lambda_c a_{bb}\sim 0.75$, ${\cal P}$ can reduce to zero with a negative ${\cal E}$, as marked by red arrow in Fig.\ref{fig1}, which gives a droplet solution satisfying condition (i). During this process, ${\cal P}_f$ and ${\cal P}_{bf}^{(2)}$ also decrease, while    
the reduction of total ${\cal P}$  mainly comes from ${\cal P}_f$, since ${\cal P}_{bf}^{(2)}$ varies in a relatively smaller scale.  
 
The suppressed Fermi pressure (${\cal P}_f$) by Rashba SOC can be attributed to the U(1) ground state degeneracy %at $k_{\perp}=\lambda$ and $k_z=0$. Such a  $U(1)$ degeneracy manifold greatly 
and thus the enhanced density-of-state $\rho(E)$ at low $E$, which approaches a constant ($\sim m_f\lambda$) as in an effective 2D geometry, rather than zero as in the usual 3D case.  % behaving like in an effective 2D geometry. %instead of zero as $E$ approaches the threshold energy. 
As a result, in the presence of SOC, more fermions can be accommodated in the low-$E$ space and for a given $n_f$ this greatly suppresses the total energy and the Fermi pressure. 
Specifically, in the low-density or strong-SOC regime where only the lower helicity branch is occupied, i.e., $n_f<n_{f,c}\equiv \lambda^3/4\pi$, we have $\lambda_f=\sqrt{4\pi n_f/\lambda}$ and 
\begin{equation}
{\cal E}_f={\cal P}_f=\frac{\pi}{\lambda m_f}  n_f^2.  \label{P_rashba}
\end{equation}
This shows that the presence of Rashba SOC can fundamentally alter the energy(pressure)-density scaling relation, from $\sim n_f^{5/3}$ in the usual case, to $\sim n_f^2$. This greatly suppresses ${\cal E}$ and ${\cal P}$ for a dilute Fermi gas (with small $n_f$). Moreover, Eq.\ref{P_rashba} shows that ${\cal E}$, ${\cal P}$ can be further reduced by increasing SOC strength $\lambda$, as also shown in Fig.\ref{fig1}.

\begin{figure}
\includegraphics[width=8.3cm, height=8.2cm] {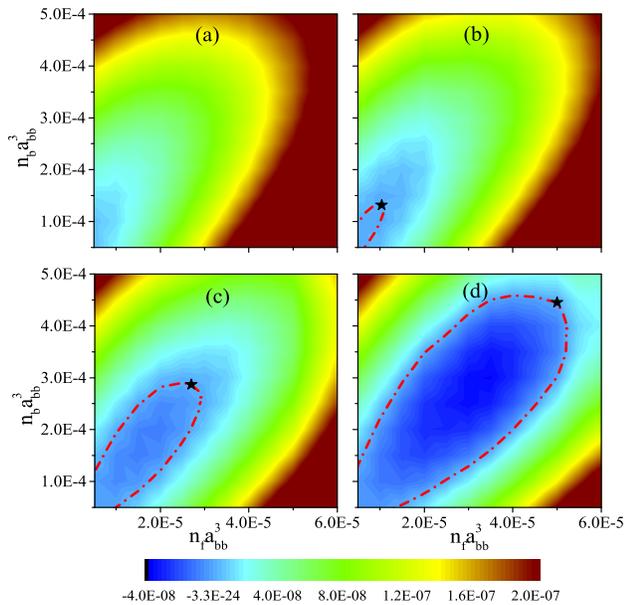}
\caption{Contour plots of ${\cal P}$ [in unit of $m_ba_{bb}^5/(2\pi V)$] in the ($n_fa_{bb}^3, n_ba_{bb}^3$) plane for different SOC strengths: $\lambda a_{bb}=0.02(a),\ 0.06(b),\ 0.08(c),\ 0.1(d)$. The red dashed-dot lines in (b,c,d) denote zero-pressure loops, and the black stars mark the locations of ground state droplets satisfying condition (ii). Here we take $a_{bf}=-3a_{bb}$ and $m_b/m_f=133/6$.}\label{P_n}
\end{figure}

Given the robust single-particle physics modified by Rashba SOC, the suppression of ${\cal P}$ should generally apply to all boson/fermion densities. In Fig.\ref{P_n}, we show the contour plots of ${\cal P}(n_b,n_f)$ for Cs-Li system taking a fixed $a_{bf}=-3a_{bb}$ and several different values of $\lambda a_{bb}$. At $\lambda a_{bb}=0.02$(Fig.\ref{P_n}(a)), ${\cal P}$ is always positive, while it can be effectively reduced when increasing $\lambda a_{bb}$ to $0.06$(Fig.\ref{P_n}(b)), where it touches zero along a small loop in $(n_b,n_f)$ plane and becomes negative inside. Further increasing $\lambda a_{bb}$ to $0.08$ and $0.1$ (Fig.\ref{P_n}(c,d)), ${\cal P}$ is further reduced and the zero-pressure loop becomes even enlarged. On each loop in (b-d), the location of ground state droplet  following condition (ii)  is further marked by a black star, and we have checked that all these solutions are with ${\cal E}<0$. In addition, because the stars all locate at the top right corner of the loops, we have $\partial {\cal P}/\partial n_{b} >0$ and $\partial {\cal P}/\partial n_{f} >0$, %and therefore by straightforward algebra the condition (iii) (Eq.\ref{stable}) is satisfied.  
which automatically ensure the satisfaction of condition (iii). Therefore they represent stable ground-state droplets satisfying all conditions (i-iii).

\begin{figure}
\includegraphics[width=8cm] {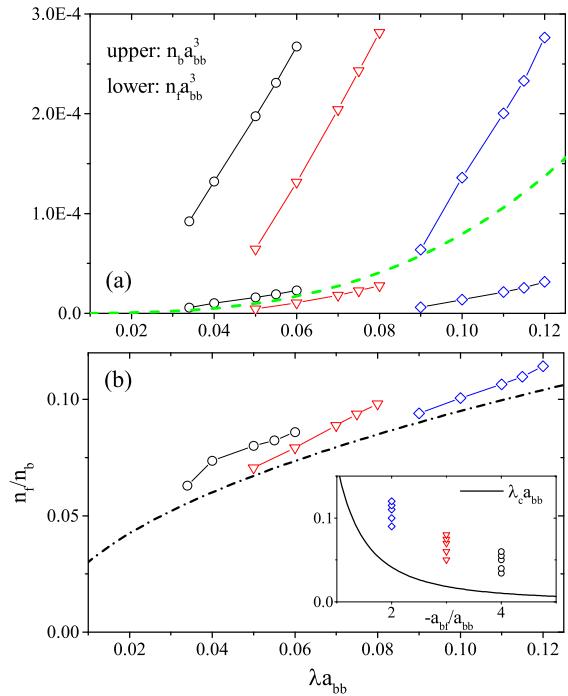}
\caption{Boson and fermion densities (a) and their ratios (b) as functions of $\lambda a_{bb}$ for the ground state droplets at different scattering lengths $a_{bf}/a_{bb}=-4$(black circles), $-3$(red triangles) and $-2$(blue diamonds). The green dashed line in (a) shows the critical fermion density $n_{f,c}=\lambda^3/(4\pi)$, below which only the lower helicity branch is occupied. In (b), the orange dashed-dot line shows fit to Eq.\ref{ratio}; in the inset, the line shows critical $\lambda_c$ for mean-field collapse (see Eq.\ref{soc_c}), and the points shows actual $\lambda$ for the data shown in the main plot. Here $m_b/m_f=133/6$.}\label{droplet}
\end{figure}

Repeating the same procedure for different attraction strengths $a_{bf}/a_{bb}=-2,\ -3,\ -4$, we show in Fig.\ref{droplet} the boson/fermion densities and their ratios as functions of $\lambda$ for the ground state Cs-Li  droplets. As shown in Fig.\ref{droplet}(a), 
%for a weaker attraction (smaller $|a_{bf}|$), stronger SOC (larger $\lambda$) is required to form the droplet, such that the Fermi pressure can be sufficiently suppressed with large $\lambda$ (as inferred by Eq.\ref{P_rashba} to make the system self-bound. F
for $a_{bf}=-4a_{bb}$(black circles), the droplets start to form at small $\lambda a_{bb}\sim 0.03$ with fermions occupying both the lower and upper helicity branches (i.e.,$n_f>n_{f,c}\equiv \lambda^3/4\pi$),  thus these droplets are mainly interaction-induced similar to those without SOC\cite{droplet_BF}. Gradually reducing attraction to $a_{bf}=-3a_{bb}$(red triangles), the droplets move to larger $\lambda$ and $n_f$ starts to drop below $n_{f,c}$. For small attraction $a_{bf}=-2a_{bb}$(blue diamonds), the droplet appears at $\lambda a_{bb}\ge 0.09$ with $n_f\ll n_{f,c}$, i.e., the fermions are located near the bottom of lower helicity branch with U(1) ground state degeneracy. Such droplet formation crucially relies on the suppressed Fermi pressure by Rashba SOC (see Eq.\ref{P_rashba}), and can only appear for strong SOC and weak attractions. Thus we term it as the SOC-induced droplet, in order to distinguish from the interaction-induced ones at small or zero SOC. Below we will extract several unique features for such kind of droplet.
%, during which process   . This is because under weak attraction, a stronger SOC is required to suppress the Fermi pressure sufficiently to make the system self-bound, as inferred by Eq.\ref{P_rashba}.   For very small 
%As the occurrence of such Bose-Fermi droplet crucially relies on the low-energy physics created by Rashba SOC, we term it as the SOC-induced droplet. Thus, by decreasing the boson-fermion attractions, the droplets complete a crossover from the interaction-induced to the SOC-induced regime.

First, given the fermion energy in Eq.\ref{P_rashba}, we see that SOC can conveniently tune the mean-field stability, and a mean-field collapse occurs for sufficiently large $\lambda$ at 
\begin{equation}
\lambda>\lambda_c= \frac{8m_bm_f}{(m_b+m_f)^2} \frac{a_{bb}}{a_{bf}^2} . \label{soc_c}
\end{equation}
The dependence of $\lambda_c$ on $a_{bf}$ is plotted in the inset of Fig.\ref{droplet}(b), and this qualitatively explains why a stronger SOC is required for droplet formation at weaker attractions, as shown in Fig.\ref{droplet}(a). %It turns out that all droplet solutions in Fig.\ref{droplet} are in the mean-field collapsing regime with $\lambda>\lambda_c$. 
Secondly, by requiring a minimal mean-field energy like in Bose-Bose mixtures\cite{Petrov}, we obtain an optimal boson/fermion density ratio as
\begin{equation}
\left(\frac{n_f}{n_b}\right)_{\rm op}= \sqrt{\frac{2m_f}{m_b}} \sqrt{\lambda a_{bb}}.\label{ratio}
\end{equation}
This shows a universal dependence of boson/fermion density ratio  on the SOC strength, which is one of characteristic features of the SOC-induced droplet.  We see from Fig.\ref{droplet}(b) that Eq.\ref{ratio} can well predict the actual density ratio for the SOC-induced droplets at $a_{bf}=-2a_{bb}$ (with small discrepancy attributed to quantum fluctuation effect), but deviate largely from that of the interaction-induced ones at stronger attraction $a_{bf}=-4a_{bb}$.

%behaves likes the mean-field  energy of bosons with an effective coupling $g\sim (\lambda m_f)^{-1}$. Omitting the quantum fluctuations, the energy of Bose-Fermi mixtures  $\sim {\cal E}_b+{\cal E}_f+{\cal E}_{bf}^{(1)}$ shares similar form to the mean-field energy of two-species bosons. Following similar analyses as in the boson-boson droplet\cite{Petrov}, we arrive at a critical $\lambda_c$ and an optimized density ratio $(n_f/n_b)_{\rm op}$ for the droplet formation in the present case:

%Eq.\ref{soc_c} explicitly explains why a larger $\lambda_c$ is required for a weaker attraction. Eq.\ref{ratio} predicts a universal function of the density ratio $n_f/n_b$ as varying $\lambda$, which does not depend on the interaction strength $a_{bf}$. In Fig.\ref{droplet}(b), we show the density ratio $n_f/n_b$ as functions of $\lambda$ for the droplet solutions in Fig.\ref{droplet}(a). We can see that Eq.\ref{ratio} can well capture the tendency of % a qualitative description to the tendency of 
%$n_f/n_b$ as varying $\lambda$, and it can even provide a quantitative prediction to the case of weak $a_{bf}=-2a_{bb}$ , where the droplet is SOC-induced instead of interaction-induced. In this case, the residue small deviations can be attributed to the presence of quantum fluctuations, i.e., the finite  ${\cal E}_{bf}^{(2)}$ and ${\cal P}_{bf}^{(2)}$, which are expected play more roles at larger $\lambda$ with larger $n_{f,b}$ (see Fig.\ref{droplet}(a)). We note that all the droplet solutions are for $\lambda>\lambda_c$, as shown in the inset of Fig.\ref{droplet}(b).

\begin{figure}[h]
\includegraphics[width=8.5cm] {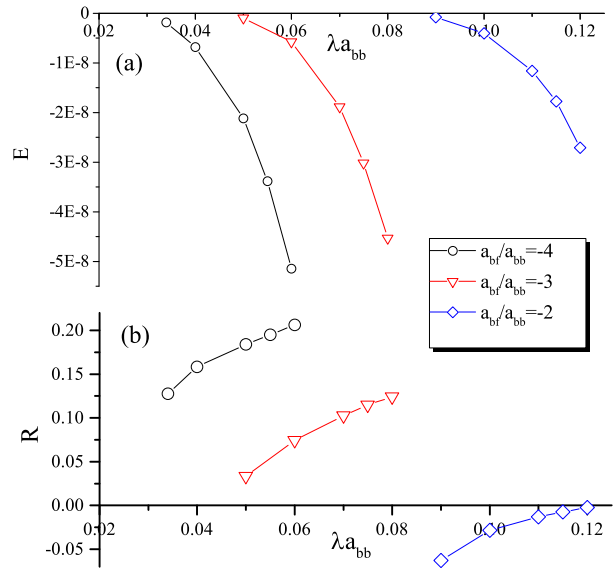}
\caption{Energy density ${\cal E}$ [in unit of  $m_ba_{bb}^5/(2\pi V)$] (a) and the ratio $R\equiv {\cal E}_{bf}^{(2)}/|{\cal E}_{bf}^{(1)}|$ (b) as functions of $\lambda$ for the droplet solutions in Fig.\ref{droplet}.}\label{droplet_E}
\end{figure}

The SOC-induced droplets also significantly differ from the interaction-induced ones in quantum fluctuations. %, quantum fluctuation can also produce dramatically different effects to the energetics of SOC-induced droplets, in comparison to that of interaction-induced ones. 
To compare the fluctuation effects for all sets of parameters,  we investigate the  relative ratio $R\equiv {\cal E}_{bf}^{(2)}/|{\cal E}_{bf}^{(1)}|$, which quantity can also be used to judge the validity of second-order perturbation theory. In Fig.\ref{droplet_E}, we plot the energy ${\cal E}$ and associated ratio $R$ for the droplet solutions in Fig.\ref{droplet}. One can see that at given $a_{bf}$, ${\cal E}$ monotonically decrease as increasing $\lambda$. For the same window of  $|{\cal E}|\in[2\times 10^{-9},3\times 10^{-8}]$, $R$ can range within $[12\%,20\%]$ for $a_{bf}=-4a_{bb}$, while can be reduced to $[3\%,10\%]$ for a weaker attraction $a_{bf}=-3a_{bb}$. For the SOC-induced droplet at $a_{bf}=-2a_{bb}$, $R$ turns negative and its absolute value can be even smaller ($\in[0,5\%]$). 

Three remarks are in order. First, the fact that the SOC-induced droplets host sufficiently small  $|R|$ is associated with their appearance in the weak coupling regime, i.e., small $a_{bf}$, which guarantees the validity of perturbative treatment in this problem as well as the practical stability in experiments. In comparison, the interaction-induced Bose-Fermi droplets have much higher $R$\cite{footnote}. Secondly, the SOC-induced droplets can exhibit a negative fluctuation energy ${\cal E}_{bf}^{(2)}<0$, which is very rare in 3D systems. This is, again, attributed to the enhanced DoS by Rashba SOC, such that more fermions can be excited near the ground-state manifold with small excitation energy and the second term in Eq. \ref{Ebf2} can dominate to produce a negative ${\cal E}_{bf}^{(2)}$. Similar effect produced by Rashba SOC have been shown to enhance the quantum depletion of a 3D Bose condensate\cite{Gordon, CZ}. Finally, despite of a negative ${\cal E}_{bf}^{(2)}$, the fluctuation pressure ${\cal P}_{bf}^{(2)}$ still keeps positive given $\partial {\cal E}_{bf}^{(2)}/\partial n_f>0$. This is why such droplet can be stabilized in the mean-field collapsing regime with $\lambda>\lambda_c$, see the inset of Fig.\ref{droplet}(b).

%We have also checked that these small attractions are not enough for the formation of universal Borromean trimers as found previously in the same setup with Rashba SOC\cite{Cui_Yi}.

%In addition, Fig.\ref{droplet_E}(b) shows that during the crossover from the interaction-induced($a_{bf}=-4a_{bb}$) to SOC-induced($a_{bf}=-2a_{bb}$) droplets,  the second-order energy (${\cal E}_{bf}^{(2)}$) can tune from positive to negative. This, again, can be attributed to the U(1) ground state degeneracy and the enhanced DoS with Rashba SOC, such that more fermions can be excited near the ground-state manifold with small excitation energy $\Delta E_{{\bf k,q}; \sigma,\sigma'}$. Therefore, the second integration term in Eq. \ref{F} can dominate over the first one, giving a negative $F$ and thus a negative ${\cal E}_{bf}^{(2)}$. We remark that such a negative energy correction from quantum fluctuations are very rare in 3D, because of the renormalization of the coupling strength. Here it is purely due to the presence of Rashba SOC, which effectively reduce the fermion dimension to 2D geometry. We note that for boson system with Rashba SOC, the same quantum fluctuation effect can lead to enhanced quantum depletions of BEC in 3D\cite{Gordon, CZ}.

Now we discuss the accessibility of SOC-induced Bose-Fermi droplets in ultracold $^{133}$Cs-$^{6}$Li and $^{87}$Rb-$^{40}$K mixtures. For a laser-generated SOC, typically the maximum $\lambda$ is given by the wave vector of two counter-propagating lasers, i.e., $\lambda_{\rm max}\sim 2\pi/1000\ {\rm nm}^{-1}$. For Cs-Li mixtures near $892$G Feshbach resonance\cite{Chin}, Cs-Cs scattering length $a_{bb}\sim 15$nm, giving $\lambda_{\rm max}a_{bb}\sim 0.1$. According to Fig.\ref{droplet}, a Cs-Li droplet can form at $a_{bf}=-2 a_{bb}=-30$nm, with densities $n_f=0.1n_b=4\times 10^{12}{\rm cm}^{-3}$. For Rb-K mixtures near $546$G resonance\cite{JILA}, given a smaller Rb-Rb scattering  length $a_{bb}\sim 5$nm, we have $\lambda_{\rm max}a_{bb}\sim 0.033$. According to Eqs.(\ref{soc_c},\ref{ratio}), %for the SOC-induced droplet at $n_f<\lambda^3/(4\pi)=2\times 10^{13}{\rm cm}^{-3}$, 
the minimum attraction required for mean-field collapse is $a_{bf}\sim-7.6a_{bb}\sim 40$nm, and the optimal density ratio in the droplet is $n_f/n_b\sim 0.17$. We note that $n_b$ is much larger than $n_f$ in both Cs-Li and Rb-K droplets, similar to Bose polaron systems as have been realized in cold atoms without SOC\cite{JILA,Aarhus,Germany}.

In summary, we have demonstrated the formation of Bose-Fermi droplets in the weak coupling regime driven by a Rashba SOC between the spin states of fermions. The SOC-induced droplets feature a negative fluctuation energy, while occur in the mean-field collapsing regime with a positive fluctuation pressure, distinct from the interaction-induced ones studied in literature. Moreover, the boson/fermion density ratio universally depends on the strength of Rashba SOC, which property can be detected in experiments. Such SOC-induced droplet has a number of implications as below. First, it offers an ideal platform to study the topological edge states when further combine SOC with interactions, since the droplet configuration naturally provides surface/boundary for cold atoms without resorting to external potentials. Moreover, this work reveals the importance of single-particle physics in engineering quantum droplets, and the associated mechanism can be generalized to a vast class of heteronuclear atomic systems in various geometries. In particular, this work sheds light on the droplet formation even in Fermi-Fermi mixtures and in mixed dimensions.

\acknowledgments
We thank D. Petrov for stimulating discussions on droplets.  This work is supported by the National Key Research and Development Program of China (2016YFA0300603), and the National Natural Science Foundation of China (No.11622436, 
No.11421092, No.11534014).


\begin{thebibliography}{99}
\bibitem{footnote} We note that the minimal $R$ can be as large as $27\%$ for Cs-Li droplets without SOC\cite{droplet_BF}.
\end{thebibliography}
\end{document}